\documentclass[letterpaper,prb,twocolumn]{revtex4} 
\usepackage{amsmath,amssymb,graphicx,times,setspace}
\usepackage{hyperref}
\begin{document}
\title{How short-range attractions impact the structural order, \\
  self-diffusivity, and viscosity of a fluid}
\author{William Krekelberg}
\email{krekel@che.utexas.edu}
\affiliation{Department of Chemical Engineering, University of Texas at Austin, Austin, TX 78712.}
\author{Jeetain Mittal}
\email{jeetain@helix.nih.gov}
\affiliation{Laboratory of Chemical Physics, NIDDK, National
  Institutes of Health, Bethesda, MD 20892-0520. }
\author{Venkat Ganesan}              
\email{venkat@che.utexas.edu}
\affiliation{Department of Chemical Engineering and Institute for
  Theoretical Chemistry, University of Texas at Austin, Austin, TX 78712.}
\author{Thomas M. Truskett}
\email{truskett@che.utexas.edu}
\thanks{Corresponding Author}
\affiliation{Department of Chemical Engineering and Institute for
  Theoretical Chemistry, University of Texas at Austin, Austin, TX 78712.}
\def\bR{ {\bf R} }
\def\br{ {\bf r}}
\def\bv{ {\bf v}}
\def\bfor{{\bf F}}
\def\bom {{\boldsymbol{\omega}}}
\def\calf{{\cal {F}}}
\def\calz{{\cal{Z}}}
\def\calg{{\cal {G}}}
\def\bstau{\boldsymbol{\tau}}
\def\rhohat{\hat{\rho}}
\def\shat{\hat{\bf S}}
\def\bu{{\bf{u}}}
\def\bM{{\bf{M}}}
\def\qdag{{q}^\dagger}
\def\mand {\text{ and }}
\def\Pc{{\phi_c}}
\newcommand{\vect}[1]{\mathbf{#1}} 
\newcommand{\RM }[1]{\mathrm{#1}}
\newcommand{\ave}[1]{ {\langle {#1} \rangle} }
\newcommand{\pd }[2]{ {\frac{\partial {#1}}{\partial {#2} } } }
\def\kB{ k_{\RM{B}} }

\begin{abstract}
  We present molecular simulation data for viscosity,
  self-diffusivity, and the local structural ordering of (i) a
  hard-sphere fluid and (ii) a square-well fluid with short-range
  attractions.  The latter fluid exhibits a region of dynamic
  anomalies in its phase diagram, where its mobility increases upon
  isochoric cooling, which is found to be a subset of a larger region of
  structural anomalies, in which its pair correlations strengthen upon
  isochoric heating.  This ``cascade of anomalies'' qualitatively
  resembles that found in recent simulations of liquid water.  The
  results for the hard-sphere and square-well systems also show that the
  breakdown of the Stokes-Einstein relation upon supercooling occurs
  for conditions where viscosity and self-diffusivity develop
  different couplings to the degree of pairwise structural ordering
  of the liquid.  We discuss how these couplings reflect dynamic
  heterogeneities.  Finally, we note that the simulation data suggests
  how repulsive and attractive glasses may generally be characterized
  by two distinct levels of short-range structural order.
\end{abstract}
\maketitle

\section{Introduction}
\label{sec:introduction}

Colloidal fluids play a central role both in technology and in the
study of condensed matter.  Regarding the latter, suspensions of
colloids are interesting model systems because they can behave
collectively in ways that are similar to atomic and molecular liquids
while simultaneously being both large and slow enough to allow
experimental measurements of their real-space structure and dynamics.
Despite the similarities between the behaviors of colloids, atoms, and
molecules, there also are important, qualitative
differences.\cite{Russel1989Colloidal-Dispe,Sciortino2002One-liquid-two-} For example, the
effective interparticle attractions between colloids can be ``tuned''
to be short-ranged relative to the particle diameter $\sigma$, unlike
the dispersion attractions in atomic and molecular systems that decay
more gradually with interparticle separation $r$, i.e., as
$(\sigma/r)^{-6}$.

Short-ranged attractive (SRA) interactions have nontrivial
implications for the static structural, equilibrium thermodynamic, and
dynamic behaviors of colloidal suspensions.  For instance, whereas
cooling a simple atomic liquid generally slows down its dynamic
processes, several recent studies have demonstrated that reducing the
temperature (or increasing the attractive interactions) of an SRA
fluid can have a non-monotonic effect on its mobility.  In fact,
concentrated SRA fluids can vitrify not only upon cooling, forming an
``attractive'' glass or gel, but also upon heating, forming a
``repulsive'' or hard-sphere (HS)
glass.\cite{Eckert2002Re-entrant-Glas,Pham2002Multiple-Glassy,Bergenholtz1999Nonergodicity-t,Fabbian1999Ideal-glass-gla,Dawson2001Higher-order-gl,Zaccarelli2002Confirmation-of,Sciortino2002One-liquid-two-}
Since the structural and mechanical properties of these glassy states
can be quite different, there is a broad interest in understanding the
properties of the precursor supercooled fluids from which they are
formed.

In this article, we explore whether the unusual effect that
temperature has on the dynamics of SRA fluids reflects a more general
connection between the static structure and the dynamics of condensed
phases.  In doing so, we find it instructive to view the behavior of
SRA fluids in the context of another well-studied system with
anomalous dynamical trends, liquid water.  Cold water behaves
differently than simple fluids in that its mobility {\em{increases}}
upon isothermal compression over a broad range of conditions.
Interestingly, the state points where water shows this behavior is a
subset of a larger region on its phase diagram where its local
structural order anomalously {\em decreases} upon
compression.\cite{Errington2001Relationship-be,Sastry2001Water-structure}
In other words, there is a ``cascade of
anomalies''\cite{Errington2001Relationship-be} where the
pressure-induced disordering of liquid water emerges at a lower
density, and ultimately disappears at a higher density, when compared
to the pressure-induced increase in its mobility.  It has been
recently
argued\cite{Mittal2006Quantitative-Li,Errington2006Excess-entropy-,Sharma2006Entropy-diffusi}
that this behavior follows from the fact that water approximately
obeys a scaling relationship between its self-diffusivity and the
structural order parameter $-s_2/\kB$, where $s_2$ is the contribution
to the fluid's excess entropy due to its static oxygen-oxygen pair
correlations, and $\kB$ is the Boltzmann constant.

Here, we use molecular dynamics simulations to systematically
investigate the relationships between static structural order
($-s_2/\kB$), self-diffusivity ($D$), and viscosity ($\eta$) for both
a HS fluid and an SRA fluid.  One of our main aims is to understand
whether the latter shows a cascade of anomalies similar to that of
liquid water.  Specifically, we are interested in whether the region
on the phase diagram of an SRA fluid where it becomes more mobile
(higher $D$, lower $\eta$) upon cooling is a subset of a larger set of
conditions where the fluid becomes more ordered (higher $-s_2/\kB$)
upon heating.  A second, and related, goal of this study is to explore
whether the relationships between $D$, $\eta$, and $-s_2/\kB$ can
provide generic insights into the breakdown of the Stokes-Einstein
(SE) relation ($D \eta / T \approx \text{constant}$) in deeply
supercooled liquids and also into the structural properties of
repulsive and attractive glasses.

The organization of the paper is as follows.  In
Section~\ref{sec:modeling-simulation}, we describe the two model
systems examined in this study and also the simulation methods used to
carry out the investigation.  Then, in
Section~\ref{sec:results-discussion}, we present our simulation
results and discuss their relevance for understanding the connection
between structural order and mobility in HS and SRA fluids.  Finally,
in Section~\ref{sec-Conclusions}, we present some concluding remarks.

\section{Modeling and Simulation}
\label{sec:modeling-simulation}

We focus on two model systems: a fluid of HS particles and a
fluid of square-well particles with short-range attractions, the
latter of which we denote the SW-SRA fluid.  The interparticle
potential ${\mathcal{V}}(r_{12})$ between particles 1 and 2 for these
systems is given by
\begin{equation}
  \label{eq:potential}
 {\mathcal{V}}(r_{12})=\begin{cases}
    \infty& r_{12}\leq\sigma_{12}, \\
    -\epsilon    & \sigma_{12}< r_{12}< \lambda \sigma_{12}, \\
    0    & r_{12}\geq \lambda\sigma_{12}. 
  \end{cases}
\end{equation}
where $r_{12}$ is the distance between the particle centers,
$\sigma_{12}=(\sigma_1+\sigma_2)/2$ is the interparticle diameter, and
the parameters $\epsilon$ and $\lambda$ set the magnitude and range of
the interparticle attraction, respectively.  For the HS fluid, one has
$\lambda=1$, and thus there are no interparticle attractions.  We
choose $\lambda=1.03$ for the SW-SRA fluid, a range similar to that of
other SW-SRA fluids known to exhibit anomalous dynamical
behavior\cite{Foffi2002Evidence-for-an,Zaccarelli2002Confirmation-of}.
In order to avoid crystallization in our study, and thus allow study
of the supercooled fluid states, we have drawn the individual particle
diameters $\sigma_i$ of both systems from a Gaussian distribution with
an average of $\sigma$ and standard deviation of $s=0.1\sigma$. For
practical reasons, we truncated this distribution so that all particle
diameters lie in the range $\sigma-3s\leq\sigma_i\leq \sigma+3s$. We
have implicitly non-dimensionalized all reported quantities in this
investigation by appropriate combinations of the characteristic length
scale, $l_{\mathrm c}=\sigma$ and a model-dependent time scale
$\tau_{\mathrm c}$.  For the HS fluid, the characteristic time scale
$\tau_{\mathrm c}$ is cast in terms of the temperature $\tau_{\mathrm
  c}=\sqrt{m \sigma^2/\kB T}$, where $m$ is the particle mass, while
for the SW-SRA fluid it is defined in terms of the attractive strength
of the potential, $\tau_{\mathrm c}=\sqrt{m \sigma^2/\epsilon}$.

To explore the behavior of these model systems, molecular dynamics
simulations were performed using an event-driven algorithm
\cite{Rapaport2004The-Art-of-Mole} in the microcanonical ensemble.
For all runs, $N=1000$ particles were simulated in a cubic simulation
cell of volume $V$ with periodic boundary conditions.  Particle
packing fractions $\Pc=\pi\sum{\sigma_i^3}/(6V)$ in the range $0.35
\le \Pc \le 0.6$ were investigated for both model systems, and
temperatures $T$ in the range $0.3 \le T \le 2.0$ were examined for
the SW-SRA fluid.  For each thermodynamic state point, between three
and ten independent simulations were performed in order to estimate
the errors in the transport coefficients.  Self-diffusivities $D$ were
calculated by fitting the long time ($t \gg 1$) mean-squared
displacements to the Einstein relation for self-diffusivity $\ave{
  \delta r^2}=6Dt$.  Zero-shear viscosities were calculated using the
impulse limit of the Einstein formula.\cite{Alder1970Studies-in-Mole}
The structural order parameter $-s_2$ was computed using the
expression
\begin{equation}
  \label{eq:s2_def}
  -s_2=\frac{\rho}{2} \int \{g(\br) \mathrm{ln} g(\br)
  -[g(\br)-1]\} d\br,
\end{equation}
where $\rho=N/V$ is the number density, and $g(\br)$ is the average
pair correlation function.  Note that $-s_2=0$ for an ideal gas, and
$-s_2\rightarrow\infty$ for a perfect crystalline lattice. Thus, as
has been discussed at length elsewhere
\cite{Truskett2000Towards-a-quant,Mittal2006Quantitative-Li,Errington2006Excess-entropy-},
one can view $-s_2$ as a scalar measure of the amount of pair-wise
translational order of the system.

\section{Results and Discussion}
\label{sec:results-discussion}

\subsection{Transport and Structural Properties}
\label{sec:transp-struct-prop}

We now examine some of the state-dependent transport and structural
properties of the HS and SW-SRA fluids.  For the HS fluid, the
quantities $D$, $\eta$, and $-s_2$ (non-dimensionalized as outlined in
Section~\ref{sec:modeling-simulation}) depend only on the packing
fraction $\Pc$, whereas the corresponding dimensionless properties for
the SW-SRA fluid depend on both packing fraction $\Pc$ and temperature
$T$.

\begin{figure}[b]
  \centering
\includegraphics{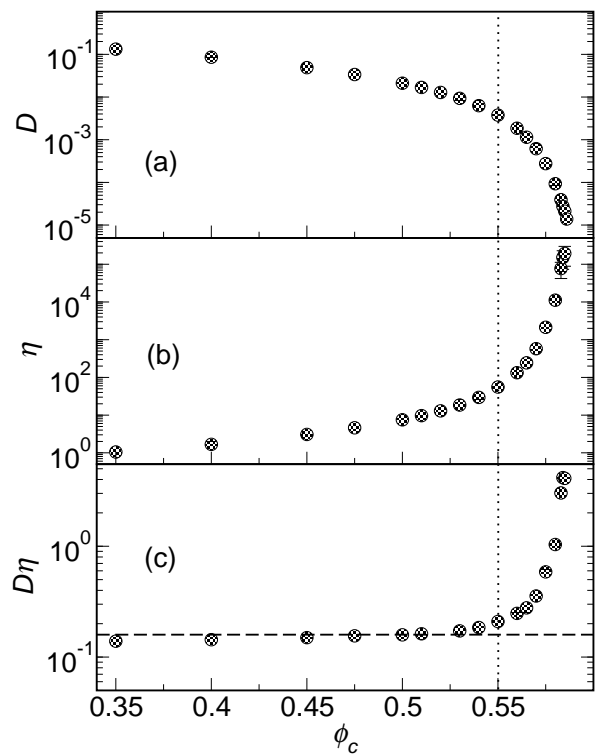}
\caption{Transport properties of the HS fluid described in the text as
  a function of packing fraction $\Pc$: (a) self-diffusivity $D$, (b)
  viscosity $\eta$, and (c) SE relationship $D\eta/T$.  The horizontal
  dashed line in (c) indicates $(2\pi)^{-1}$, the expected value of
  the SE relation in the slip limit, and the vertical line denotes the
  point at which which $D\eta=1.2/(2\pi)$.  In this work, we use
  this simple heuristic to identify the breakdown of the SE relation.} 
  \label{fig:HS_transport}
\end{figure}

The transport properties of the HS fluid are displayed in
Fig.~\ref{fig:HS_transport}.  As expected, with increasing packing
fraction $\Pc$, the self-diffusivity $D$ monotonically decreases,
while the viscosity $\eta$ monotonically increases.  The changes in
both of these properties become more pronounced for $\Pc > 0.55$,
conditions for which $D(\Pc)$ can be accurately fitted using either
Vogel-Fulcher or power-law functional
forms~\cite{Kumar2006Nature-of-the-b} with a predicted divergence near
$\Pc\sim0.6$.

Figure~\ref{fig:HS_transport}(c) illustrates that the SE relation in
the slip limit, $D \eta \approx (2 \pi)^{-1}$, is approximately obeyed
by this HS fluid for $\Pc<0.55$.  However, for higher packing
fractions, large positive deviations from the slip limit of the SE
relation become apparent.~\cite{Kumar2006Nature-of-the-b} This type of
SE ``breakdown'' has been observed in studies of several glass-forming
fluids in their deeply supercooled
states.\cite{Rossler1990Indications-for,Cicerone1995How-do-molecule,Cicerone1993Photobleaching-,Fujara1992Translational-a}

We note that, as observed by previous
investigators,\cite{Kumar2006Nature-of-the-b} the SE relation is not
strictly obeyed by the HS fluid even at lower packing fractions.  For
example, the product $D \eta$ varies by approximately $30\%$ as the
packing fraction is increased from $\Pc=0.35$ to $0.55$.  However, the
breakdown of the SE relation is typically understood to occur when the
product begins to exhibit a pronounced positive deviation from the
slip value.  In this work, we identify the breakdown of the SE
relation by the heuristic criterion, $D\eta \ge1.2/(2\pi)$.  As can be
seen from Figure~\ref{fig:HS_transport}(c), this threshold is crossed
at $\Pc\approx0.55$ for the HS fluid investigated here.

\begin{figure}[t]
  \centering
  \includegraphics{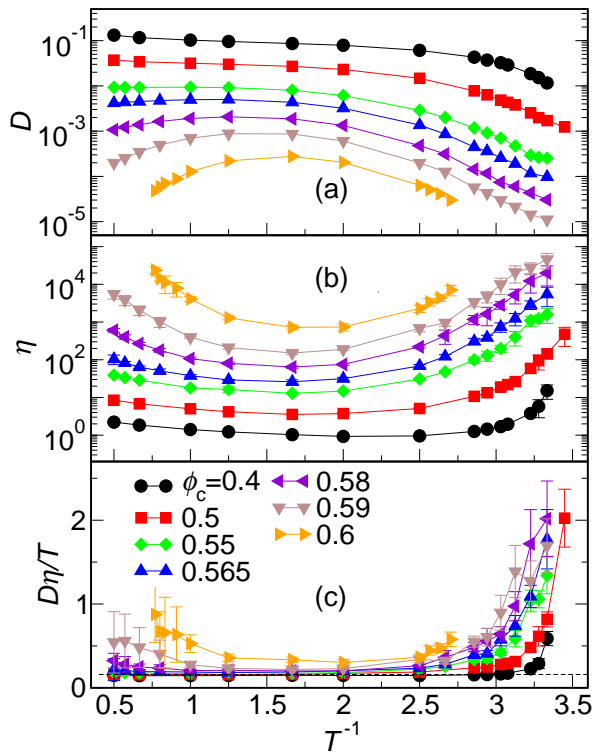}
  \caption{ Transport properties of the SW-SRA fluid described in the
    text as a function of packing fraction $\Pc$ and reciprocal
    temperature$T^{-1}$: (a)
    self-diffusivity $D$, (b) viscosity $\eta$, and (c) SE
    relationship $D\eta/T$.  The horizontal dashed line in (c)
    indicates $(2\pi)^{-1}$, the expected value of the SE relation in
    the slip limit.} 
\label{fig:sw_trans_vs_temp}
\end{figure}

The simulated transport properties of the SW-SRA fluid are displayed
in Fig.~\ref{fig:sw_trans_vs_temp} as a function of reciprocal
temperature $T^{-1}$ along isochores.  The most striking feature of
this plot is that $D$ exhibits a maxima with inverse temperature, a
behavior that has also been observed in both experiments and computer
simulations of other SRA
fluids.~\cite{Eckert2002Re-entrant-Glas,Pham2002Multiple-Glassy,Bergenholtz1999Nonergodicity-t,Fabbian1999Ideal-glass-gla,Dawson2001Higher-order-gl,Zaccarelli2002Confirmation-of,Puertas2003Simulation-stud,Puertas2002Comparative-Sim,Puertas2004Dynamical-heter}
As discussed in Section~\ref{sec:introduction}, this trend becomes
pronounced at high $\Pc$, where the system can ultimately form either
a repulsive glass by isochoric heating or an attractive glass or gel
via isochoric cooling.

The behavior of the zero-shear viscosity $\eta$ of the SW-SRA fluid,
displayed in Fig.~\ref{fig:sw_trans_vs_temp}(b), qualitatively mirrors
that of its self-diffusivity.  For a more quantitative comparison, the
SE relationship $D \eta /T$ is plotted in
Figure~\ref{fig:sw_trans_vs_temp}(c).  Note that the slip limit of the
SE relation is again approximately obeyed for the SW-SRA fluid over a
broad range of $T^{-1}$ and $\Pc$.  However, for all values of $\Pc$
investigated here, the SE relation breaks down at sufficiently low
$T$, as the attractive glass transition is approached.  At high $T$,
on the other hand, only the highest $\Pc$ isochore studied showed a
significant breakdown of the SE relation.  This asymmetry between high
and low $T$ behaviors has also been observed in other model SRA
fluids.\cite{Puertas2003Simulation-stud,Puertas2002Comparative-Sim,Foffi2002Evidence-for-an,Zaccarelli2002Confirmation-of} It
simply reflects the fact that one must reach a relatively high value
of $\Pc$ in these systems to achieve the level of frustration required
to form a repulsive glass, whereas interparticle attractions can
induce formation of the attractive glass at much lower particle
concentrations.

\begin{figure}[!h]
  \centering
  \includegraphics{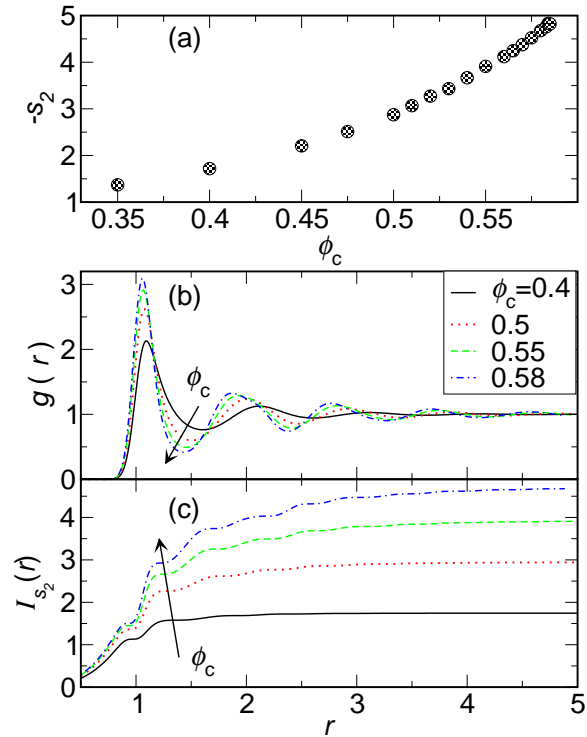}
  \caption{Structural properties of the HS fluid discussed in the
    text. (a) Translational structural
    order parameter $-s_2$ versus packing fraction $\Pc$.  (b) Radial
    distribution function $g(r)$ for several packing fractions. (c)
    Cumulative order integral (see Eq.~\ref{eq:Is2}).  Arrows
    indicate increasing $\Pc$.} 
  \label{fig:HS_s2}
\end{figure}

It is natural to wonder whether the decrease in mobility that the HS
and SW-SRA fluids experience upon approach to the glass transition is
generally correlated with an increase in the amount of local
structural order that they exhibit.  To explore this issue, we first
examine the behavior of the structural order parameter $-s_2$ for the
HS fluid.  As can be seen in Fig.~\ref{fig:HS_s2}(a), $-s_2$ for this
system monotonically increases with packing fraction, indicating a
strengthening of the pair-wise interparticle correlations.  From the
radial distribution functions $g(r)$ shown in Fig.~\ref{fig:HS_s2}(b),
it is also evident that these correlations correspond to the
progressive development of well-defined coordination shells around the
particles.  The contributions of these shells to the translational
order parameter $-s_2$ become readily apparent when we investigate the
following cumulative order integral,
\begin{equation}
  \label{eq:Is2}
  I_{s_2}(r)=2\pi \rho\int_0^r r'^{2} \{g(r') \RM{ln}g(r') - [g(r')-1]\} dr'. 
\end{equation}
Note that one recovers $-s_2$ from this integral in the large $r$
limit, and thus $I_{s_2}(r)$ quantifies the average amount of
translational ordering on length scales smaller than $r$ surrounding a
particle.  In Fig.~\ref{fig:HS_s2}(c), we observe nearly step-wise
increases in $I_{s_2}(r)$ at the locations of the various coordination
shells.  It is also clear that increasing $\Pc$ of the HS fluid has
two main effects on its structural order.  It strengthens the ordering
within the individual coordination shells, and it uniformly increases
the number of coordination shells (i.e., the range of order).

\begin{figure}[!h]
  \centering
  \includegraphics{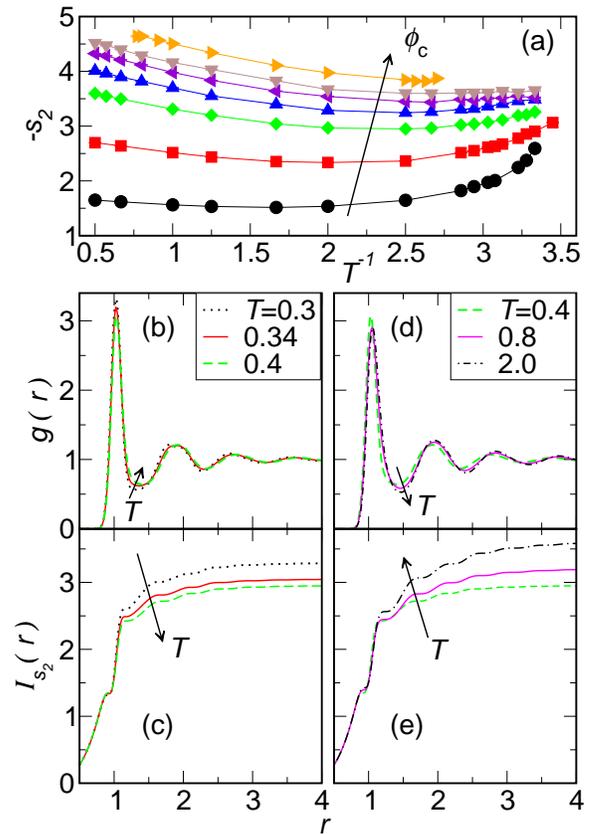}
  \caption{Structural properties of the SW-SRA fluid. (a)
    Structural order parameter $-s_2$ versus reciprocal temperature~$T^{-1}$. 
    Symbols are the same as in Fig.~\ref{fig:sw_trans_vs_temp}. (b)
    Radial distribution function $g(r)$ and (c) cumulative
    order integral~$I_{s_2}(r)$ for $\Pc=0.55$ and
    $T \le 0.4$. (d) Radial distribution function $g(r)$ and (e)
    cumulative order integral~$I_{s_2}(r)$ for $\Pc=0.55$ and
    $T \ge 0.4$.  }
  \label{fig:SW_s2}
\end{figure}

The behavior of the structural order parameter $-s_2$ for the SW-SRA
fluid is displayed in Fig.~\ref{fig:SW_s2}(a) as a function of
reciprocal temperature $T^{-1}$ along isochores.  In accord with what
might be expected based on the behavior of this fluid's transport
coefficients, $-s_2(T)$ displays broad minima.  At low $T$, heating
the fluid decreases its structural order (similar to what happens in
normal molecular fluids), but, at high $T$, heating anomalously
increases its structural order.

To determine the origins of this trend, we focus on the behaviors of
the radial distribution function and the cumulative order integral for
the SW-SRA fluid along the $\Pc=0.55$ isochore.  As can be seen in
Fig.~\ref{fig:SW_s2}(b)-(e), heating induces subtle changes in $g(r)$
that result in nontrivial cumulative changes in the structural order.
In particular, heating the cold fluid ($T \le 0.4$) results in a
barely detectable decrease in the height of the first peak of $g(r)$,
and it has little effect on the other coordination shells.  However,
as is shown in Fig.~\ref{fig:SW_s2}(c), even these subtle
modifications to $g(r)$ result in significant changes to the
cumulative order integral of the fluid.  In particular, heating the
cold fluid decreases the total amount of pair-wise structural order,
even though the range of the order remains essentially the same.  This
change is qualitatively consistent with the thermal weakening of
transient clustered configurations of physically ``bonded'' particles
observed in SRA fluids at lower
temperatures.\cite{Krekelberg2006Model-for-the-f,Krekelberg2006Free-Volumes-an,Puertas2004Dynamical-heter,Sciortino2002One-liquid-two-}

The effect of increasing temperature on the structure of the warm
SW-SRA fluid ($T>0.4$) is different [see Fig.~\ref{fig:SW_s2}(d) and
(e)].  In particular, heating continues to decrease the height of the
first peak in $g(r)$, but it also broadens the peak and enhances the
interparticle correlations associated with the other, more distant,
coordination shells.  The net effect is an {\em increase} in both the
total amount of translational structural order and its range.  This
structural change is due to the fact that heating the warm fluid
collapses the open channels of free volume that form at intermediate
$T$ due to weak interparticle
clustering,\cite{Krekelberg2006Free-Volumes-an} essentially jamming
the particles into a less efficient, and more correlated, packing
arrangement.\cite{Pham2002Multiple-Glassy,Sciortino2002One-liquid-two-,Zaccarelli2002Confirmation-of,Krekelberg2006Free-Volumes-an}

It is clear from the results presented in this section that there is a
qualitative (negative) correlation between pair-wise structural order
and mobility in both the HS and SW-SRA fluids.  In the following two
sections, we explore the extent to which this connection can be made
quantitative.

\subsection{Connection between structure and mobility anomalies}
\label{sec:conn-betw-anom}

As discussed in Section~\ref{sec:transp-struct-prop}, at high $T$ and
$\Pc$, both the self-diffusivity and the structural order of the
SW-SRA fluid behave in a manner that is anomalous when compared to
simple molecular fluids [see Figs.~\ref{fig:sw_trans_vs_temp}(a) and
\ref{fig:SW_s2}(a)].  Whereas increasing $T$ of a simple fluid
generally increases its mobility, isochorically heating the warm
SW-SRA fluid can result in slower single-particle dynamics.  This
latter behavior is characterized by the following inequality:
\begin{equation}
  \label{eq:diffusive_anom}
  \biggl(\pd{D}{T}\biggr)_{\Pc}<0, \quad \text{self-diffusivity anomaly}. 
\end{equation}
The structural order of a simple molecular fluid, on the other hand,
normally decreases when it is isochorically heated.  Therefore, we
denote conditions for which the following inequality holds (i.e.,
order increases upon heating),
\begin{equation}
  \label{eq:structural_anom}
  \biggl(\pd{[-s_2]}{T}\biggr)_{\Pc}>0, \quad  \text{structural anomaly}, 
\end{equation}
as ``structurally anomalous''.

\begin{figure}[!h]
  \centering
  \includegraphics[width=3in]{SW_anomalies-scaled}
  \includegraphics[width=3.1in]{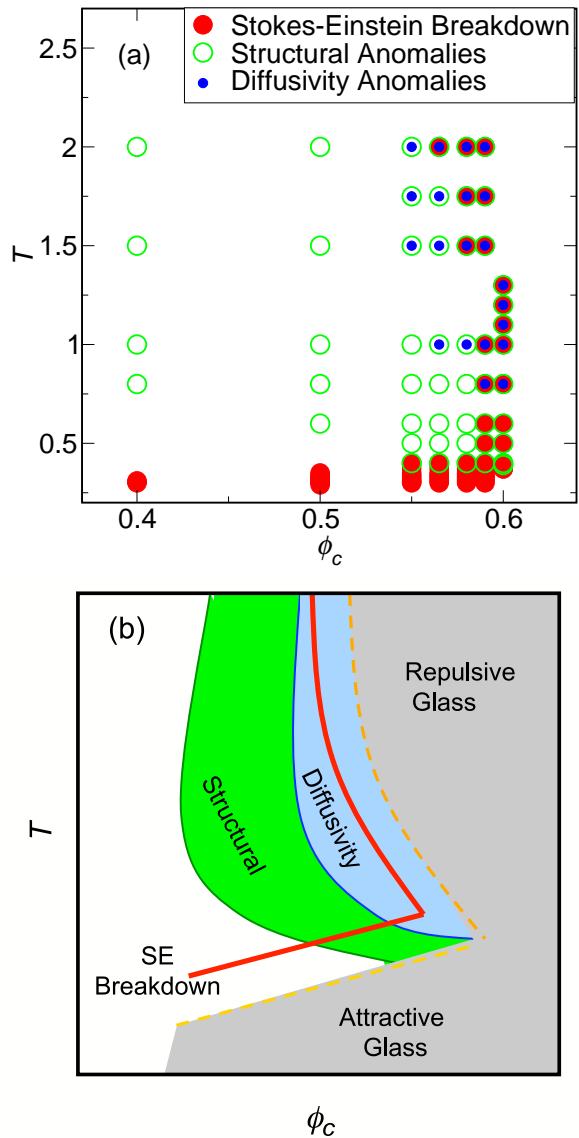}
  \caption{Conditions exhibiting self-diffusivity and structural
    anomalies, as well as breakdown of the SE relation 
[$D\eta/T>1.2/(2\pi)$, see discussion in
    text] for the SW-SRA fluid in the $T$-$\Pc$ plane.  (a) Results
    from simulations.  Large closed circles are state points where the
    SE relationship breaks down.  Open circles
    represent the region of structural anomalies defined by
    Eq.~\eqref{eq:structural_anom}.  Small closed circles represent
    the region of self-diffusivity anomalies defined by
    Eq.~\eqref{eq:diffusive_anom}. (b) Schematic representation of the
    data.  The green shaded region (and area to its right)
    represents state points where the fluid is structurally anomalous.  
The blue shaded
  region represents state points exhibiting the self-diffusivity anomaly. 
   Points to the right of the red curve show a breakdown of the SE
   relation. The gray region represents the repulsive and attractive
   glassy states.} 
  \label{fig:anomalies_schematic}
\end{figure}

The locations of the regions for self-diffusivity and structural
anomalies of the SW-SRA fluid in the $T$-$\Pc$ plane, as determined by
numerical differentiation of the data in
Figs.~\ref{fig:sw_trans_vs_temp}(a) and \ref{fig:SW_s2}(a),
respectively, are displayed in Fig.~\ref{fig:anomalies_schematic}(a).
A schematic based on the data is provided in
Fig.~\ref{fig:anomalies_schematic}(b).  The most striking point is
that the region of structural anomalies appears to completely envelop
the region of self-diffusivity anomalies.  In other words, the SW-SRA
fluid exhibits unusual $T$-dependencies for its single-particle
dynamics only for those state points where it also exibits unusual
$T$-dependencies for its structural order.

This ``cascade'' of structural and dynamic anomalies shown by the
SW-SRA fluid is very similar to that observed originally in
simulations of the SPC/E model of
water~\cite{Errington2001Relationship-be} and later in simulations of
other simpler models that also show waterlike behavior.~\cite{Yan2006Family-of-tunab,Oliveira2006Structural-anom,Oliveira2006Thermodynamic-a,Errington2006Excess-entropy-,Shell2002Molecular-struc}  Recall that cold water
behaves differently from simple fluids over a wide range of conditions
in that its mobility increases, while its structural order decreases,
when it is isothermally compressed.  Thus, the generic similarity
between the behavior of water and the SW-SRA fluid shown in
Fig.~\ref{fig:anomalies_schematic} is that, in both cases, the region
on the phase diagram where mobility anomalies occur is a subset of the
region where structural anomalies are found.  It has been recently
argued\cite{Mittal2006Quantitative-Li,Errington2006Excess-entropy-,Sharma2006Entropy-diffusi}
that this type of behavior for water follows directly from the fact
that liquid water approximately obeys a scaling relationship between
its self-diffusivity and its translational structural order parameter
$-s_2$ over a wide range of temperature and density.  In the next
section, we test whether there exist similar scaling relations between
$-s_2$ and the transport coefficients of the HS and SW-SRA fluids.  We
also discuss how such relations might provide insights into the
breakdown of the SE relation for these systems.

\subsection{Structure-property relations and the breakdown of
  Stokes-Einstein}
\label{sec:conn-betw-stok}

As alluded to above, recent molecular dynamics
simulations\cite{Mittal2006Quantitative-Li}, motivated by other
earlier observations of
Rosenfeld\cite{Rosenfeld1977Relation-betwee,Rosenfeld1999A-quasi-univers}
and Dzugutov\cite{Dzugutov1996A-univeral-scal}, have demonstrated that
the following simple relation is approximately obeyed by various model
fluids in their equilibrium liquid states:
\begin{subequations}
  \label{eq:transport_s2}
  \begin{equation}
    \label{eq:diffusion_s2}
    D=A_D\, \RM{exp}[B_D\, s_2],
  \end{equation}
  where $A_D$ and $B_D$ are parameters which may depend on packing
  fraction (i.e., density), but not on $T$.  Results from other
  earlier theoretical studies,\cite{Rosenfeld1977Relation-betwee} and
  considerations based on the Stokes-Einstein relation, suggest that a
  similar relationship should approximately hold for the zero-shear
  viscosity $\eta$ of these equilibrium fluids:
  \begin{equation}
    \label{eq:viscosity_s2}
    \eta=A_\eta \,\RM{exp}[B_\eta\, s_2]
  \end{equation}
\end{subequations}
where, again, $A_\eta$ and $B_\eta$ may depend on packing fraction
(density) only.

However, given that the SE relation breaks down as a liquid is
supercooled, it is apparent that Eq.~\eqref{eq:diffusion_s2} and
\eqref{eq:viscosity_s2}, with coefficients fit to higher temperature
equilibrium fluid data, cannot also describe the transport
coefficients in deeply supercooled liquid states.  Nonetheless, it has
recently been shown that the functional form of
Eq.~\eqref{eq:diffusion_s2} can in fact approximately describe the
isochoric self-diffusivity data of model supercooled liquids over a
broad range of temperatures if a different pair of parameters $A'_D$
and $B'_D$ are adopted.\cite{Mittal2006Quantitative-Li} In other
words, there appears to be a crossover upon cooling where the
self-diffusivities of fluids transition from being approximately
described by $D=A_D\, \RM{exp}[B_D\,s_2]$ for equilibrium states to
being approximately described by $D=A'_D\, \RM{exp}[B'_D\,s_2]$ for
supercooled conditions.  Of course, even this latter relation must
eventually fail for conditions very near the glass transition where
both $D$ and $\eta^{-1}$ rapidly vanish, while $s_2$ remains
finite.~\cite{Truskett2000Towards-a-quant} We will return to this last
point at the end of the section.

The structure-property scalings discussed above suggest several
interesting questions concerning the liquid state.  For example, does
the aforementioned crossover between scaling relations occur near the
breakdown of the SE relation?  Furthermore, does a similar crossover
for the $s_2$ dependence of $\eta$ occur upon supercooling?  If so,
does it coincide with the crossover point for $D$?  Below, we use our
molecular dynamics simulation results to investigate these questions
for the HS and SW-SRA fluids.  The idea is that the new information
that we gain about how $D$ and $\eta$ couple to pair structure should
give insights into the SE breakdown and the general effects that
supercooling has on liquids.

\begin{figure}[!h]
  \centering
  \includegraphics{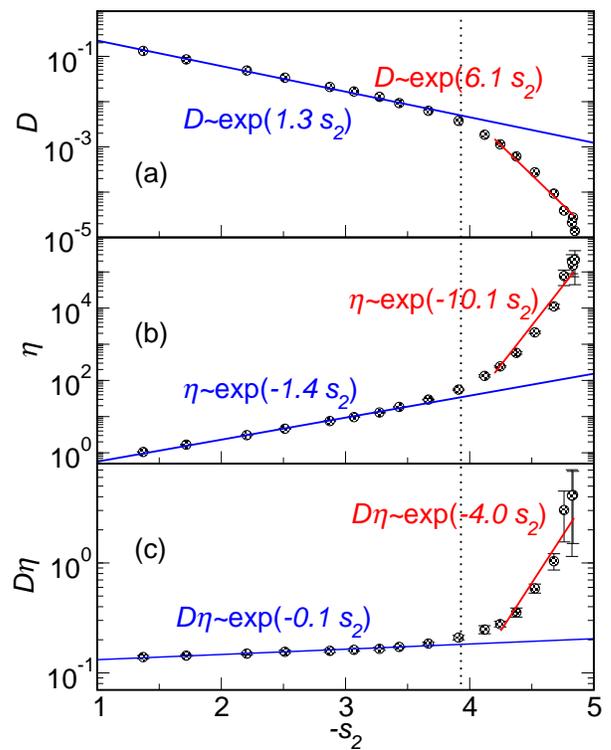}
  \caption{Transport properties as a function of
    structural order parameter $-s_2$ for the HS fluid described in
    the text. (a) Self-diffusivity $D$, (b)
    viscosity $\eta$, and (c) the SE relationship.  In (a) and
    (b), the blue and red lines represent fits to
    Eq.~\ref{eq:transport_s2} for the equilibrium and supercooled
    states, respectively.} 
  \label{fig:HS_KT_trans_vs_s2}
\end{figure}

First, we consider the HS fluid.  Figure~\ref{fig:HS_KT_trans_vs_s2}
displays the transport coefficients and the SE relationship for this
system as a function of the structural order order parameter, $-s_2$,
which itself depends only on packing fraction $\Pc$.  Although the
basic scaling relations outlined above describe the behavior of
attractive fluids along isochores, they apply equally well for
athermal fluids (such as this) along isotherms, keeping in mind that
the glass transition in the latter is approached by compression and
not by cooling.  One important aspect to note in
Fig.~\ref{fig:HS_KT_trans_vs_s2} is that there is, in fact, a
crossover between ``equilibrium'' (low $-s_2$) and ``supercooled''
(high $-s_2$) states.  In fact, Eq.~\eqref{eq:transport_s2} provides a
quantitative fit to the data for packing fractions in the range
$0.35<\Pc<0.55$ ($1.4 < -s_2 < 3.9$).  Interestingly, the point at
which both transport coefficients diverge from this equilibrium
scaling relationship closely coincides with the packing fraction
($\Pc\approx 0.55$) where $D \eta= 1.2/(2\pi)$, i.e., the breakdown of
the SE relation.

It is also evident from Fig.~\ref{fig:HS_KT_trans_vs_s2} that one can
use the scaling form of Eq.~\eqref{eq:transport_s2} with different
pairs of coefficients to approximately describe the $s_2$ dependencies
of the transport coefficients for the supercooled HS fluid.  The fits
of Eq.~\eqref{eq:transport_s2} for these supercooled states, however,
are not as accurate as those for the equilibrium fluid data below the
crossover.  In fact, our only goal in fitting the supercooled liquid
data to this exponential form is that it allows us to extract simple
quantitative measures [$B'_D$ and $B'_{\eta}$] of the couplings that
exist between the transport coefficients and the static structure of
the fluid.  It can be seen both from the raw data in
Fig.~\ref{fig:HS_KT_trans_vs_s2} and from the values of these
coefficients for the ``equilibrium'' and ``supercooled'' HS fluid
[($B_D=1.3$, $B_\eta=-1.4$) and ($B'_D=6.1$, $B'_\eta=-10.1$),
respectively] that the breakdown of the SE relation coincides with
qualitative change in how static structure correlates to $D$ and
$\eta$.  In the equilibrium fluid, both transport coefficients show
weak, and roughly equivalent, couplings to $s_2$.  However, after the
SE breakdown, the transport coefficients develop much stronger
couplings to structural order.  This is presumably due to the integral
role that cooperative structural rearrangements play in the relaxation
of deeply supercooled liquids.\cite{Debenedetti1996Metastable-Liqu}

It is clear from Fig.~\ref{fig:HS_KT_trans_vs_s2} that the reason that
the SE relation breaks down in this system is because the viscosity of
the deeply supercooled HS fluid becomes much more sensitive to changes
in static structural order than the self-diffusivity.  This is
consistent with the observations of previous studies that have
correlated the breakdown of the SE relationship to the onset of
heterogeneous
dynamics.\cite{Kumar2006Nature-of-the-b,Yamamoto1998Heterogeneous-D,Ediger2000Spatially-heter,Stillinger1994Translation-rot,Tarjus1995Breakdown-of-th}
Heterogeneous dynamics is typically characterized by the presence of
many particles that transiently exhibit exceedingly high or low values
mobility relative to the mean.  The highly mobile particles have been
shown to readily diffuse distances on the order of a particle diameter
by so-called ``hopping'' motions.  It is presumably the presence of
these highly mobile particles that allows the self-diffusivity to
maintain a weaker coupling to static structure than the viscosity.

\begin{figure}[t]
  \centering
  \includegraphics{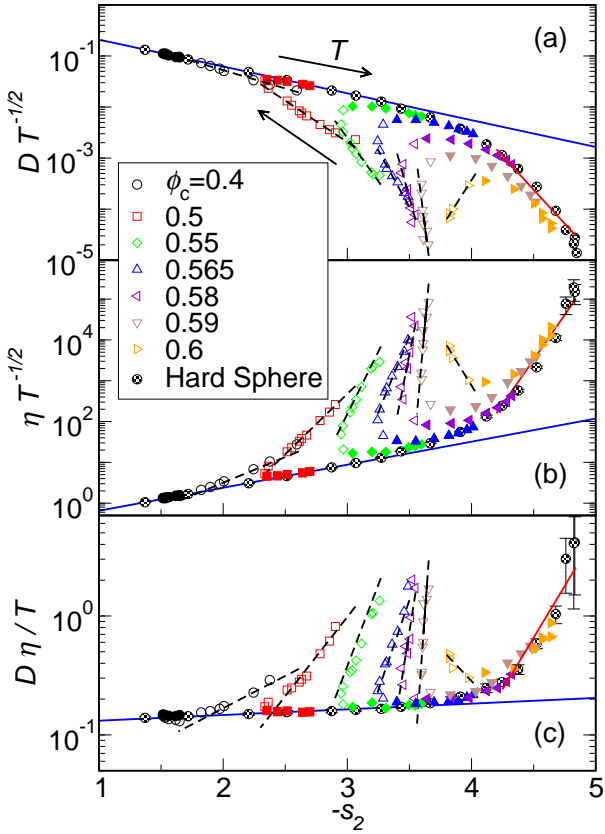}
  \caption{(a) Scaled self-diffusivity $DT^{-1/2}$, (b) 
scaled viscosity $\eta T^{-1/2}$, and (c)
  SE relationship $D\eta/T$ versus structural order
  parameter $-s_2$ for the SW-SRA fluid at several packing
  fractions $\Pc$. Self-diffusivity $D$, viscosity $\eta$, and the 
SE relationship $D\eta$ for the HS fluid are provided for comparison. Filled and open symbols
  represent the high ($T>0.5$) and low ($T<0.5$) temperature branches
  of the SW-SRA fluid, respectively.  Dashed lines in (a) and (b) are
  fits of the SW-SRA data to
  Eq.~\eqref{eq:transport_s2} and the dashed lines in (c) are the
  products of the respective fits in (a) and (b). Arrows in (a)
  indicate the general direction of increasing $T$. Red and blue lines
  have the same meaning as those in Fig.~\ref{fig:HS_KT_trans_vs_s2}.} 
\label{fig:sw_scaled_trans_vs_s2}
\end{figure}

\begin{figure}[!h]
  \centering
  \includegraphics{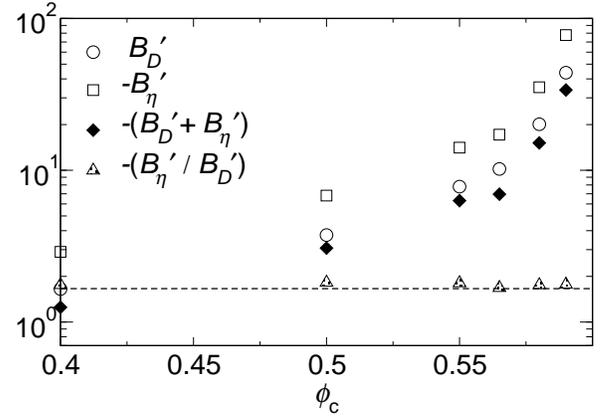}
  \caption{Values of the coupling coefficients $B'_D$ and $-B'_{\eta}$
    from fits of the supercooled SW-SRA data to
    Eq.~\eqref{eq:transport_s2} (shown in
    Fig.~\ref{fig:sw_scaled_trans_vs_s2}) for the self-diffusivity $D$
    and viscosity $\eta$, respectively.  Also shown are the (negative)
    sum $-(B'_D+B'_{\eta})$ and the ratio $-(B'_{\eta}/B'_D)$ of the
    exponents.  The dashed line represents the HS fluid value for
    $-(B'_{\eta}/B'_D)$ .} 
  \label{fig:s2_slopes_isochores}
\end{figure}

Displayed in Fig.~\ref{fig:sw_scaled_trans_vs_s2} are the scaled
transport coefficients, $DT^{-1/2}$ and $\eta T^{-1/2}$, of the SW-SRA
fluid as a function of the structural order parameter $-s_2$.  The
$T^{-1/2}$ factor is included here to remove the trivial ``thermal
velocity'' contribution to these transport coefficients, which then
allows them to be directly compared to the dimensionless values of $D$
and $\eta$ of the HS fluid.  Along these lines, it should be noted
that the high temperature data points for $DT^{-1/2}$, $\eta
T^{-1/2}$, and $D \eta /T$ of the SW-SRA fluid ($T>0.5$, denoted in
Fig.~\ref{fig:sw_scaled_trans_vs_s2} by filled symbols) are indeed
approximately described by the corresponding data for $D$, $\eta$ and
$D \eta$ of the HS fluid.  One consequence of this is that the
breakdown of the SE relation for the SW-SRA fluid upon approaching the
repulsive glass transition by heating occurs at approximately the same
value of $-s_2$ as the breakdown of the SE relation for the HS fluid
upon compression.  In contrast, the value of $-s_2$ of the SW-SRA
fluid at the breakdown of the SE relation upon cooling toward the
attractive glass is different for each packing fraction studied.  As
might be expected, fluids with lower packing fractions show departures
from the slip limit of the SE relation for conditions where they
exhibit lower amounts of translational structural order.

Do the SW-SRA transport coefficients also follow ``equilibrium'' and
``supercooled'' exponential scaling branches when plotted versus
$-s_2$?  For $\Pc<0.55$ and intermediate temperatures, the data
approximately collapse onto the same exponential scaling relation
obtained from the fit of the equilibrium HS fluid data (shown in all
panels of Fig.~\ref{fig:sw_scaled_trans_vs_s2} as a blue line).  At
sufficiently high or low temperatures, however, the SW-SRA data
transitions to ``supercooled'' exponential scalings with different
coupling coefficients $B'_D$ and $B'_\eta$.  Interestingly, these
transitions closely coincide with the breakdown of the SE relation.
In all cases, similar to the HS fluid, $D$ exhibits a considerably
weaker dependence on $-s_2$ than does $\eta$ (i.e., $-B'_\eta>B'_D$,
see Fig.~\ref{fig:s2_slopes_isochores}) after the SE breakdown.
Although the difference between the coupling coefficients increases
with $\Pc$, the relative magnitude remains approximately constant
($-B'_\eta/B'_D\sim1.7$) and very similar to that of the HS fluid.
This latter structure-property connection is general in the sense that
it approximately holds for these model liquids as they become
``supercooled'' via heating, cooling, or compression.

\begin{figure}[!h]
  \centering
   \includegraphics[width=3.1in]{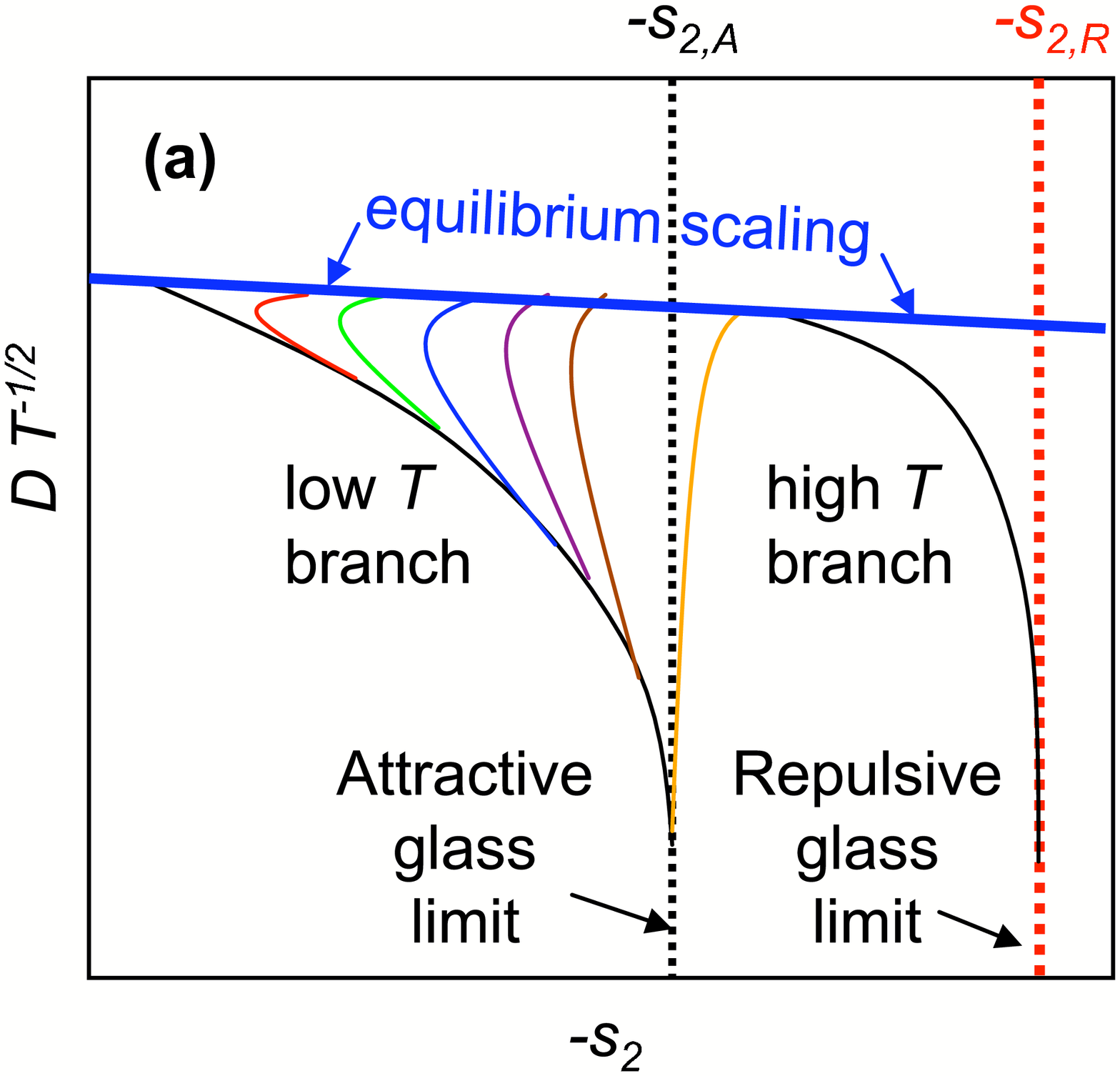}
   \includegraphics[width=3.1in]{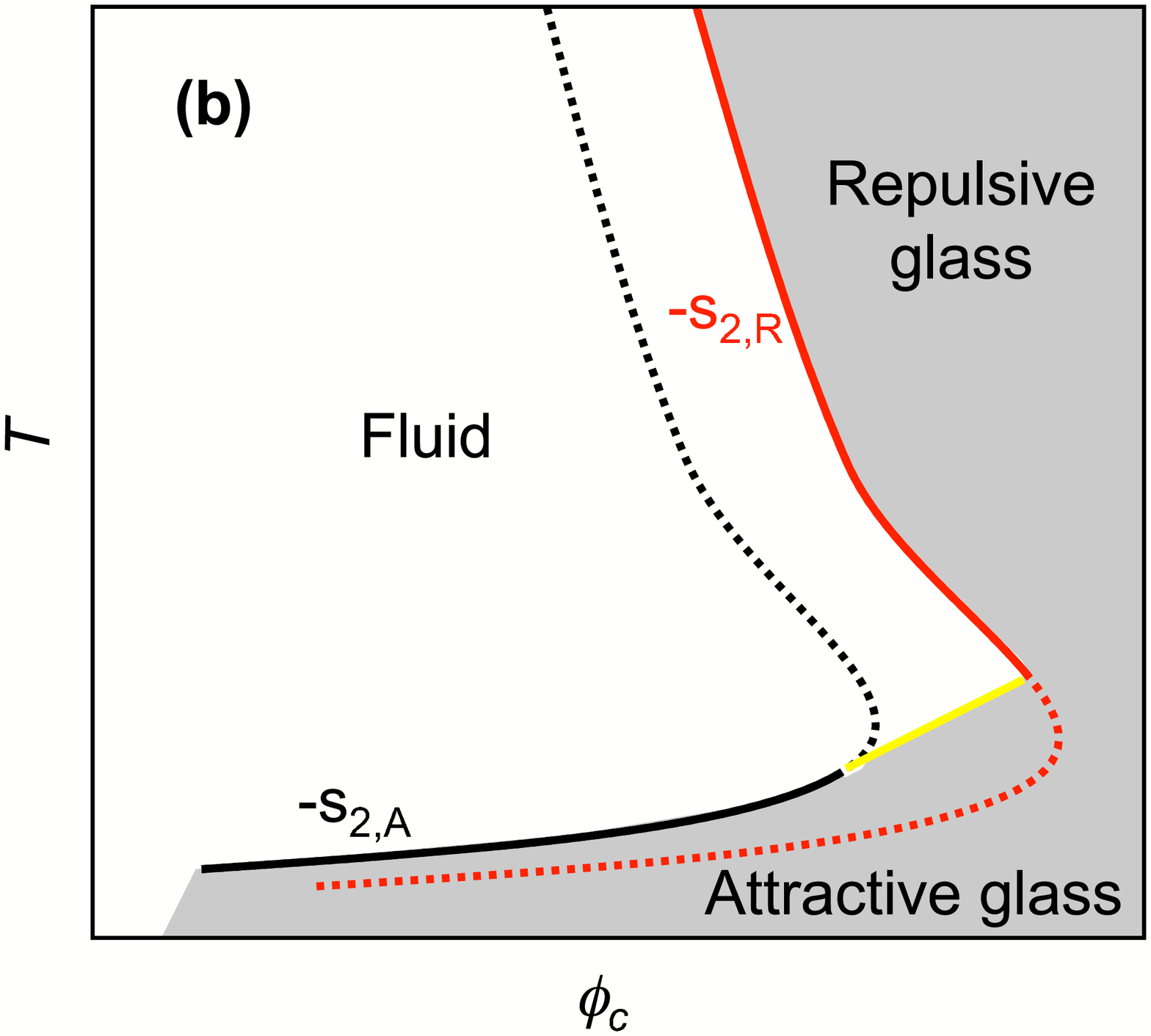}
  \caption{(a) Schematic representing a speculation about how the
    scaled self-diffusivity vs $-s_2$ isochores of the SW-SRA fluid
    [Fig.~\ref{fig:sw_scaled_trans_vs_s2}(a)] might behave as the
    repulsive and attractive glass transitions are approached.  The
    quantities $s_{2,R}$ and $s_{2,A}$ represent the limiting values
    of $s_2$ for the repulsive and attractive glasses, respectively. 
    (b) The black and red curve are the proposed iso-$s_2$ loci in the
    $T-\Pc$ plane at the charactertic repulsive and attractive glass
    values, respectively (discussed in the text).  The portions of
    these lines that are solid represent the hypothesized glass transition.  The
    yellow line is a narrow transition region where the glass line
    is proposed to cross between the iso-$s_2$ curves. }
  \label{fig:s2-glass-schematic}
\end{figure}

We would like to conclude this section with a speculation.  Although
it is not yet possible to equilibrate molecular dynamics simulations
very close to the repulsive or attractive glass transitions in the
SW-SRA fluid, the data shown in
Fig.~\ref{fig:sw_scaled_trans_vs_s2}(a) and (b) are suggestive of what
may happen to the structural order of the fluid in those limits.  As
discussed earlier, the high temperature behaviors of the SW-SRA
transport coefficients, when plotted versus $-s_2$, approximately
follow the trends of the HS fluid.  Therefore, it is reasonable to
suspect that the structural order parameter of the high temperature
(repulsive) glass $-s_{2,R}$ will be close to that of the HS glass
[Fig.~\ref{fig:s2-glass-schematic}(a)].  At low temperatures, the
transport coefficients of the SW-SRA fluid as a function of $-s_2$
also appear as if they may asymptote to a different, lower, limiting
value [Fig.~\ref{fig:s2-glass-schematic}(a)], i.e., a value of the
structural order parameter $-s_{2,A}$ characteristic of the attractive
glass.  The resulting picture is that the repulsive and attractive
glass lines closely follow two ``iso-$s_2$'' curves, with a sharp
transition between the two in a narrow temperature range, shown
schematically in Fig.~\ref{fig:s2-glass-schematic}(b)).  Since $s_2$
can be readily obtained from static pair correlations, this is a
speculation that could be tested via experiments of SRA colloidal
fluids.~\cite{Weeks2000Three-Dimension}  It would also be interesting to explore in future studies the
extent to which liquid-state (e.g., mode-coupling) theories for SRA
fluids are able to reproduce the empirical connection between $s_2$
and dynamics found in our simulations and to test the predictions that
they make about how the structural order varies along the ideal
repulsive and attractive glass lines.

\section{Conclusions}
\label{sec-Conclusions}

We have presented new molecular simulation data for viscosity,
self-diffusivity, and the local structural ordering of both a
hard-sphere fluid and a square-well fluid with short-range attractions
(relative to the particle diameter).  We found that the latter system
has a region of mobility anomalies in the temperature-packing fraction
plane, where its self-diffusivity increases upon isochoric cooling.
This region is entirely enclosed within a wider set of state points
where the fluid's pair correlations strengthen upon isochoric heating.
This type of ``cascade of anomalies'' is very similar to that found in
recent simulations of liquid water, and it follows from a broader
connection between static structure and dynamics in condensed phase
systems.

Both the hard-sphere and square-well fluids show that the breakdown of
the Stokes-Einstein relation upon supercooling occurs for conditions
where viscosity and self-diffusivity develop different couplings to
the degree of pairwise structural ordering of the liquid.  We
discussed how these couplings reflect dynamic heterogeneities.
Finally, we provided an experimentally testable hypothesis about how
repulsive and attractive glasses may be generally characterized by two
distinct levels of short-range structural order.  In future work, we
will investigate whether there are similar connections between
non-equilibrium dynamics (e.g.  shear-dependent viscosity) and
structural order in these systems.

\section*{Acknowledgments}
\label{sec:acknowledgments}

WPK acknowledges financial support of the National Science Foundation
for a Graduate Research Fellowship.  TMT acknowledges financial
support of the National Science Foundation (CTS 0448721), the David
and Lucile Packard Foundation, and the Alfred P. Sloan Foundation. VG
acknowledges financial support of the Robert A. Welch Foundation and
the Alfred P.~Sloan Foundation.  Computer simulations for this study
were performed at the Texas Advanced Computing Center (TACC).

\newpage

\end{document}